\documentstyle[12pt]{article}

\begin{document} 
\vskip 3cm
\begin{center}
{\large Effect of the additional second neighbor hopping
on the spin dynamics in the $t$-$J$ model}\\
\medskip
Xianglin Ke$^{a}$, Feng Yuan$^{b}$, and Shiping Feng$^{a,b}$ \\
$^{a}$Department of Physics, Beijing Normal University,
Beijing 100875, China\\
$^{b}$National Laboratory of Superconductivity, Academia
Sinica, Beijing 100080, China\\
\end{center}
\bigskip

The effect of the additional second neighbor hopping $t'$ on the
spin dynamics in the $t$-$J$ model is studied within the
fermion-spin theory. Although the additional second neighbor
hopping $t'$ is systematically accompanied with the reduction of
the dynamical spin structure factor and susceptibility, the
qualitative behaviors of the dynamical spin structure factor and
susceptibility are the same as in the case of $t$-$J$ model. It is
also shown that the integrated dynamical spin structure factor is
almost $t'$ independent in the underdoped regime.

\newpage

Following the initial discovery of antiferromagnetic (AF) spin
fluctuations in copper oxide materials , extensive experiments and
theoretical studies have been carried out in order to clarify the
relationship between these AF spin fluctuations and superconductivity
[1,2]. Many authors have studied the spin dynamics of copper oxide
materials based on the $t$-$J$ model, and the results are in
qualitative agreement with experiments [2]. However, the
angle-resolved photoemission spectroscopy measurements on copper
oxide materials show that the $t$-$t'$-$J$ or $t$-$t'$-$t''$-$J$
models can give a good explanation of the experimental data near
$k=(\pi,0)$ point in the Brillouin zone [3], where $t'$ and $t''$
are the second- and third-nearest neighbors hopping matrix elements,
respectively. Then a natural question is what is the effect of these
additional hoppings on the spin dynamics of the $t$-$J$ model. In
this paper, we discuss this issue within $t$-$t'$-$J$ model,
\begin{eqnarray}
H&=&-t\sum_{i\hat{\eta}\sigma}C^{\dagger}_{i\sigma}C_{i+\hat{\eta}
\sigma}+t'\sum_{i\hat{\tau}\sigma}C^{\dagger}_{i\sigma}
C_{i+\hat{\tau}\sigma} \nonumber \\
&+&J\sum_{i\hat{\eta}}{\bf S}_{i}\cdot {\bf S}_{i+\hat{\eta}},
\end{eqnarray}
where $\hat{\eta}=\pm \hat{x},\pm\hat{y}$, $\hat{\tau}=\pm \hat{x}
\pm\hat{y}$, $C^{\dagger}_{i\sigma}$ is the electron creation operator,
and ${\bf S}_{i}$ is spin operator. The $t$-$t'$-$J$ model (1) is
supplemented by the on-site local constraint $\sum_{\sigma}
C^{\dagger}_{i\sigma}C_{i\sigma}\leq 1$, ${\it i.e.}$, there are no
doubly occupied sites. This on-site local constraint can be treated
exactly in analytical calculations within the fermion-spin theory [4],
$C_{i\uparrow}=h^{\dagger}_{i}S^{-}_{i}$,
$C_{i\downarrow}=h^{\dagger}_{i}S^{+}_{i}$, where the spinless fermion
operator $h_{i}$ keeps track of the charge (holon), while the
pseudospin operator $S_{i}$ keeps track of the spin (spinon). Therefore
the fermion-spin theory naturally incorporates the physics of the
charge-spin separation. Within the fermion-spin theory, the spin
dynamics of the $t$-$J$ model has been discussed by considering
spinon fluctuations [5]. Following their discussions, we can obtain
the dynamical spin structure factor and susceptibility of the
$t$-$t'$-$J$ model as $S(k,\omega)=2(1-e^{-\beta\omega})^{-1}
{\rm Im}D(k,\omega)$ and $\chi''(k,\omega)=2{\rm Im}D(k,\omega)$,
respectively, where the full spinon Green's function $D(k,\omega)$
has been given in Ref. [6] by using the loop expansion to the
second-order.

We have performed a numerical calculation for the spin dynamics
of the $t$-$t'$-$J$ model, and the results of the dynamical structure
factor spectrum and dynamical susceptibility spectrum at AF wave vector
$Q=(\pi,\pi)$ in the doping $\delta=0.06$ for the temperature $T=0.2J$
with the parameters $t/J$=2.5, $t'/J$=0 (solid line), $t'/J$=0.5 (dashed
line), and $t'/J$=0.8 (dash-dotted line) are plotted in Fig. 1(a) and
Fig. 1(b), respectively. From these results, we find that although
the additional second neighbor hopping is systematically
accompanied with a clear reduction of the dynamical spin structure
factor and susceptibility, the qualitative property of the dynamical
spin structure factor and susceptibility is not changed. The spin
structure factor spectrum is separated into low- and high-frequency
parts, respectively, but the high-frequency part is suppressed in the
susceptibility, which is consistent with the experiments [7].
For a further understanding the effect of $t'$ on the spin dynamical of
$t$-$J$ model, we have discussed the behavior of the integrated dynamical
spin structure factor \={S}$(\omega)=(1+e^{-\beta\omega})S_{L}(\omega)$
with $S_{L}(\omega)=(1/N)\sum_{k}S(k,\omega)$, and the numerical results
in the doping $\delta=0.06$ and $\delta=0.15$ for the temperature
$T=0.2J$ with the parameters $t/J$=2.5, $t'/J$=0 (solid line), $t'/J$=0.5
(dashed line), and $t'/J$=0.8 (dash-dotted line) are shown in Fig. 2(a)
and Fig. 2(b), respectively. These results show that the integrated spin
structure factor of the $t$-$t'$-$J$ model is almost $t'$ independent
in the underdoped regime. $\-S(\omega)$ is decreased with increasing
energies for $\omega <0.5t$, and almost constant for $\omega >0.5t$,
which is also consistent with the experiments [7] and numerical
simulations [8].

\vskip 2cm
\centerline{Acknowledgements}
\vskip 1cm

This work was supported by the National Natural Science Foundation and
the State Education Department of China through the Foundation of
Doctoral Training.

\newpage
\centerline{Figures}
\vskip 1cm

FIG. 1. (a) $S(Q,\omega)$ and (b) $\chi''(Q,\omega)$ in the doping
$\delta=0.06$ for $T=0.2J$ with $t/J$=2.5, $t'/J$=0 (solid line),
$t'/J$=0.5 (dashed line), and $t'/J$=0.8 (dash-dotted line).

FIG. 2. \={S}$(\omega)$ in the doping (a) $\delta=0.06$ and (b)
$\delta=0.15$ for $T=0.2J$ with $t/J$=2.5, $t'/J$=0 (solid line),
$t'/J$=0.5 (dashed line), and $t'/J$=0.8 (dash-dotted line).

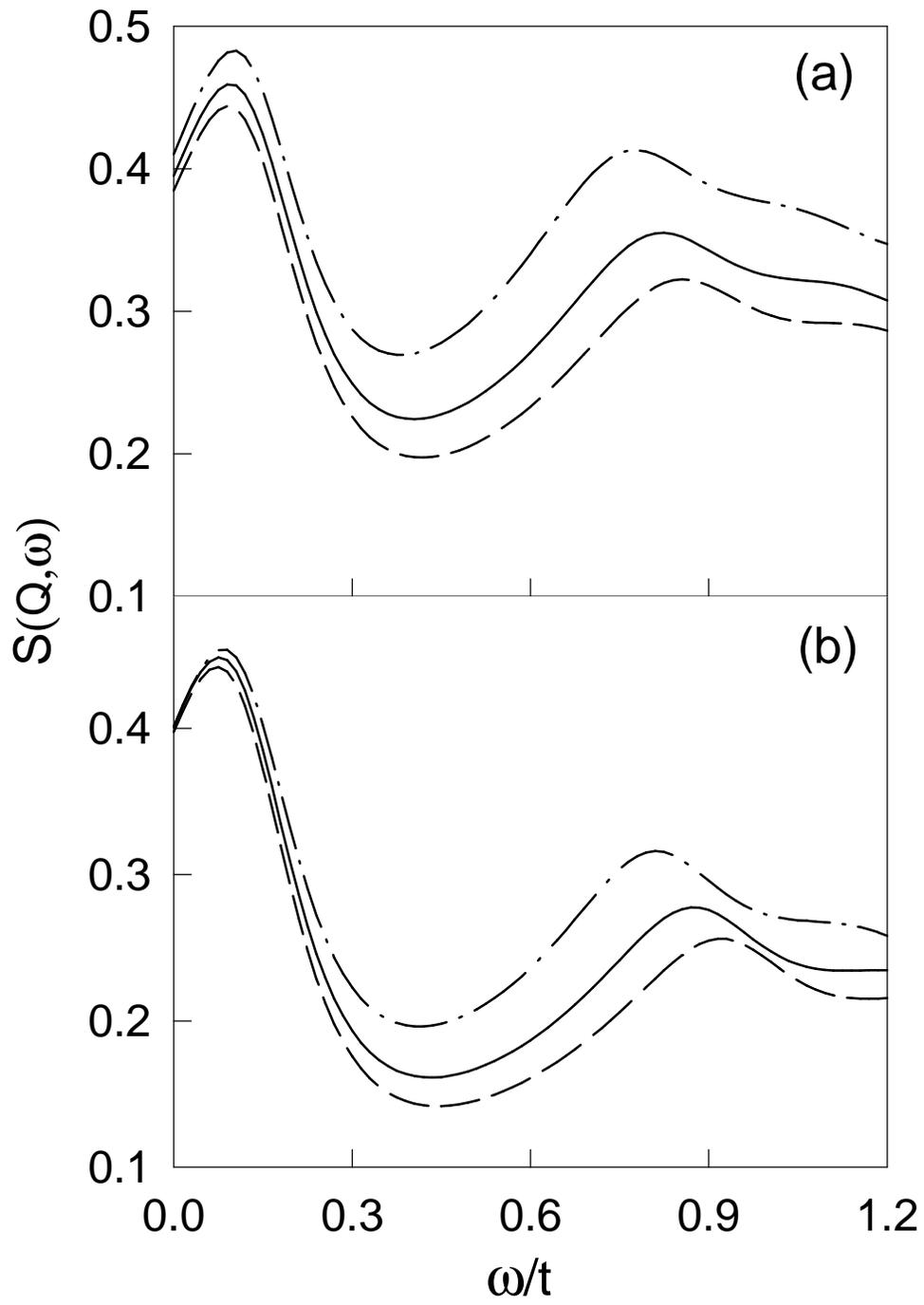
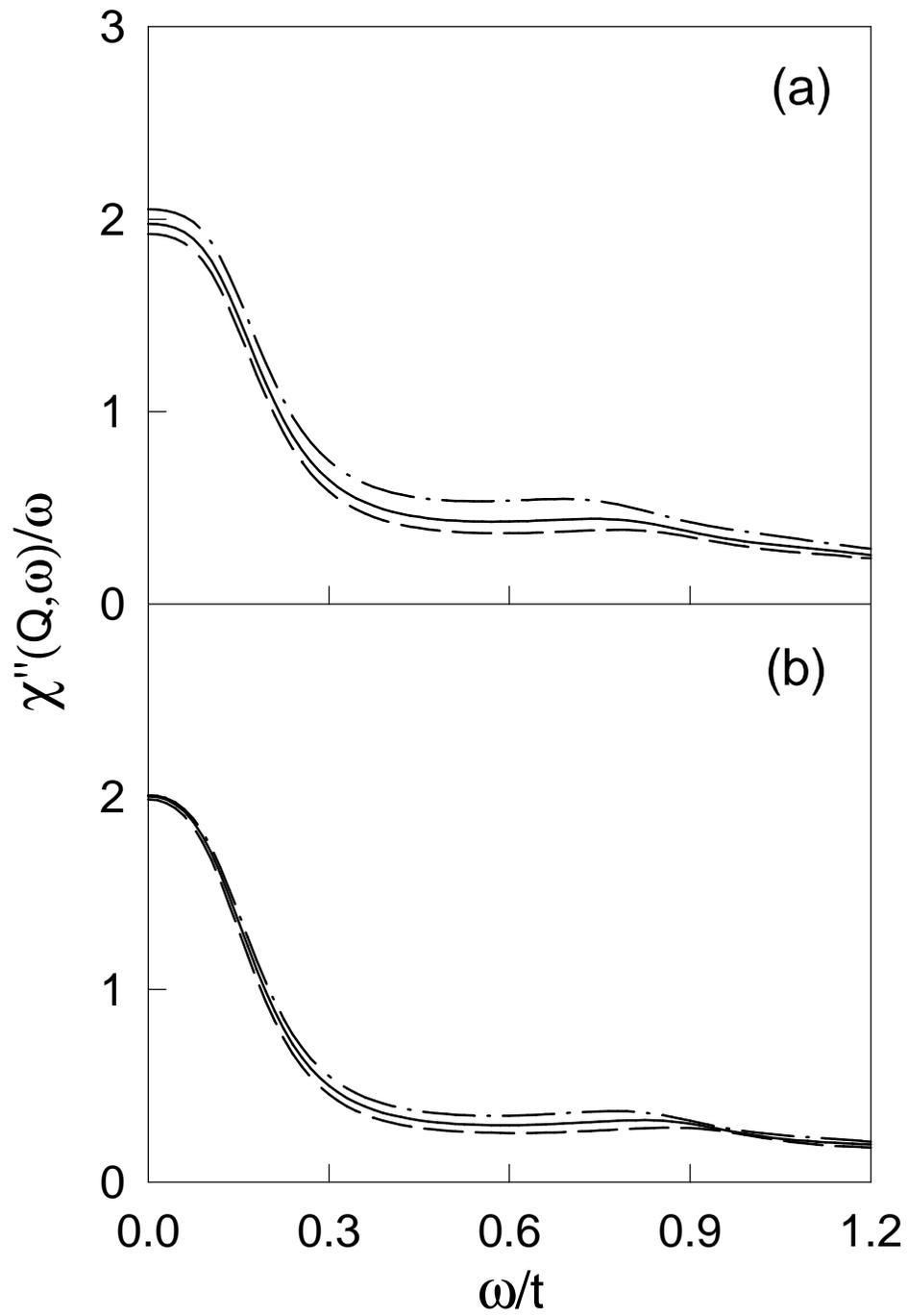
\end{document}